\documentclass{interact}

\usepackage[utf8]{inputenc}
\usepackage{hyperref}
\usepackage[numbers,sort&compress]{natbib}

\articletype{RESEARCH ARTICLE}

\title{Perturbative Analytical Framework for Thermal Wave Diffusion in Non-linear Building Envelopes}

\author{
	\name{Corentin Guigot\thanks{CONTACT Corentin Guigot. Email: cguigot@cesi.fr}}
	\affil{CESI LINEACT, Campus CESI, 24, Le Paquebot-CS 60133, Saint-Nazaire, 44600, France}
}

\begin{document}
	
	\maketitle
	
	\begin{abstract}
		Model Predictive Control (MPC) in building energy management requires transient thermal models balancing thermodynamic accuracy with computational efficiency. Standard spatial discretization triggers state-space inflation, paralyzing real-time solvers, while analytical Transfer Matrix Methods (TMM) suffer from high-frequency numerical overflow and assume material homogeneity.
		
		This paper introduces a frequency-domain framework based on the continuous spatial Riccati equation. A recursive admittance mapping strictly bounds exponential growth, preventing numerical instability. Regular perturbation theory analytically resolves continuous spatial property gradients ($\lambda(x)$) and non-linear $T^4$ radiative boundaries as equivalent harmonic source terms. 
		
		This meshless approach eliminates spatial truncation errors. It analytically corrects peak heating load deviations of 21.9\% in wetted media and mitigates artificial nocturnal cooling fluxes of 12.0~W/m$^2$. Preserving an $\mathcal{O}(N)$ spatial complexity, the framework structurally avoids state-space inflation, ensuring the high-speed execution demanded by multi-week MPC optimization.
	\end{abstract}
	
	
\section{Introduction}

The integration of buildings into smart energy grids through Model Predictive Control (MPC) requires dynamic thermal models capable of strictly bounding prediction errors over multi-day stochastic horizons \cite{drgona2020all, wang2022science, oldewurtel2012use, henze2004evaluation, judkoff1995international, kummert2001optimal}. The fundamental challenge in building performance simulation lies in the strict trade-off between thermodynamic accuracy and computational complexity \cite{hensen2012building, walton1983tarp}. 

Transient heat conduction through multi-layered envelopes is resolved either through discrete time-domain numerical methods or continuous frequency-domain analytical solutions. Discretization techniques, such as the Finite Difference Method (FDM) or Finite Element Method (FEM), provide the flexibility to model spatial heterogeneities by defining $M_s$ discrete spatial nodes. While implicit temporal schemes \cite{crank1947practical} resolve the temporal instability of explicit formulations \cite{courant1967partial}, they structurally enforce a high-dimensional state-space. Suppressing numerical diffusion requires refined spatial grids, generating an inflation of state variables ($\mathcal{O}(\sum M_s)$) that paralyzes the real-time execution of urban-scale stochastic MPC algorithms \cite{kansal2025review}.

To circumvent spatial discretization, frequency-domain spectral methods, such as the Transfer Matrix Method (TMM) \cite{davies1973thermal}, resolve the one-dimensional heat equation analytically. By mapping the envelope as a cascade of linear quadrupoles, these methods achieve a spatial execution speed of $\mathcal{O}(N)$ (where $N$ is the number of macroscopic layers). While algorithmically efficient, unmodified spectral methods suffer from two structural limitations. First, the matrix evaluation of positive exponential arguments systematically triggers 64-bit floating-point overflow \cite{goldberg1991every, higham2002accuracy} under high-frequency transient forcing. Second, the analytical propagator structurally relies on the assumption of material homogeneity. 

Materials in applied building physics are rarely purely homogeneous. Thermophysical properties, particularly the thermal conductivity of structural and porous layers ($\lambda$), exhibit a continuous spatial dependence driven by temperature gradients \cite{budaiwi2002variations} and interstitial moisture accumulation \cite{elassaad2024influence}. Standard state-space models systematically enforce a strict homogeneous spatial assumption to preserve easily invertible transmission matrices \cite{athienitis1986analytical}. This homogeneous assumption introduces a truncation error on peak conductive heating loads, often exceeding a 20\% relative deviation under severe hygrothermal conditions \cite{tariku2010transient}, which compromises the sizing of HVAC systems and the reliability of peak-shaving strategies.

To bypass the matrix-induced instability of the TMM, this paper presents a scalar frequency-domain framework based on the continuous spatial Riccati equation \cite{bellman1959invariant, pagneux1996study}. A recursive admittance mapping is derived to analytically bound positive exponential wave arguments, neutralizing high-frequency numerical overflow without requiring internal spatial discretization. 

The linear propagator is extended via regular perturbation theory \cite{bender1999advanced} to integrate continuous spatial property gradients ($\lambda(x)$) and non-linear Stefan-Boltzmann boundary conditions \cite{koch2025multi} as harmonic source terms. This formulation projects state-dependent thermophysical discrepancies directly into the macroscopic admittance update. By preserving a strictly $\mathcal{O}(N)$ spatial complexity per harmonic, the framework structurally avoids the state-space inflation inherent to discrete meshing, enabling the high-speed temporal execution ($\mathcal{O}(M \log M + N \cdot M)$) demanded by applied urban-scale MPC architectures.

\section{Formalism: Thermal Admittance and the N-Layer Propagator}

\subsection{The Harmonic Heat Equation and Thermal Admittance}

A one-dimensional multi-layered building envelope comprising $N$ nominally homogeneous isotropic strata is considered. Each layer $j \in [1, N]$ is defined by its thickness $e_j$, thermal conductivity $\lambda_j$, density $\rho_j$, and specific heat capacity $c_{p,j}$. The spatial coordinate $x$ is defined along the outward normal to the envelope, originating at the internal environment interface ($x=0$).

Under periodic boundary conditions, the time-domain temperature field $T_j(x,t)$ and conductive heat flux $\Phi_j(x,t)$ are decomposed into Fourier harmonics. For a given angular frequency $\omega$, the spatial temperature distribution $\tilde{T}_j(x)$ satisfies the one-dimensional Helmholtz equation \cite{carslaw1959conduction, corones1975bremmer}:
\begin{equation}
	\frac{d^2 \tilde{T}_j(x)}{dx^2} - q_j^2 \tilde{T}_j(x) = 0
	\label{eq:helmholtz}
\end{equation}
where $q_j$ is the complex thermal wave vector governing the dissipative propagation of the thermal wave:
\begin{equation}
	q_j = \sqrt{i\omega \frac{\rho_j c_{p,j}}{\lambda_j}} = (1+i)\sqrt{\frac{\omega}{2\alpha_j}}
	\label{eq:wavevector}
\end{equation}
with $\alpha_j = \lambda_j / (\rho_j c_{p,j})$ denoting the thermal diffusivity of the layer.

\begin{figure}[htbp]
	\centering
	\includegraphics[width=0.95\textwidth]{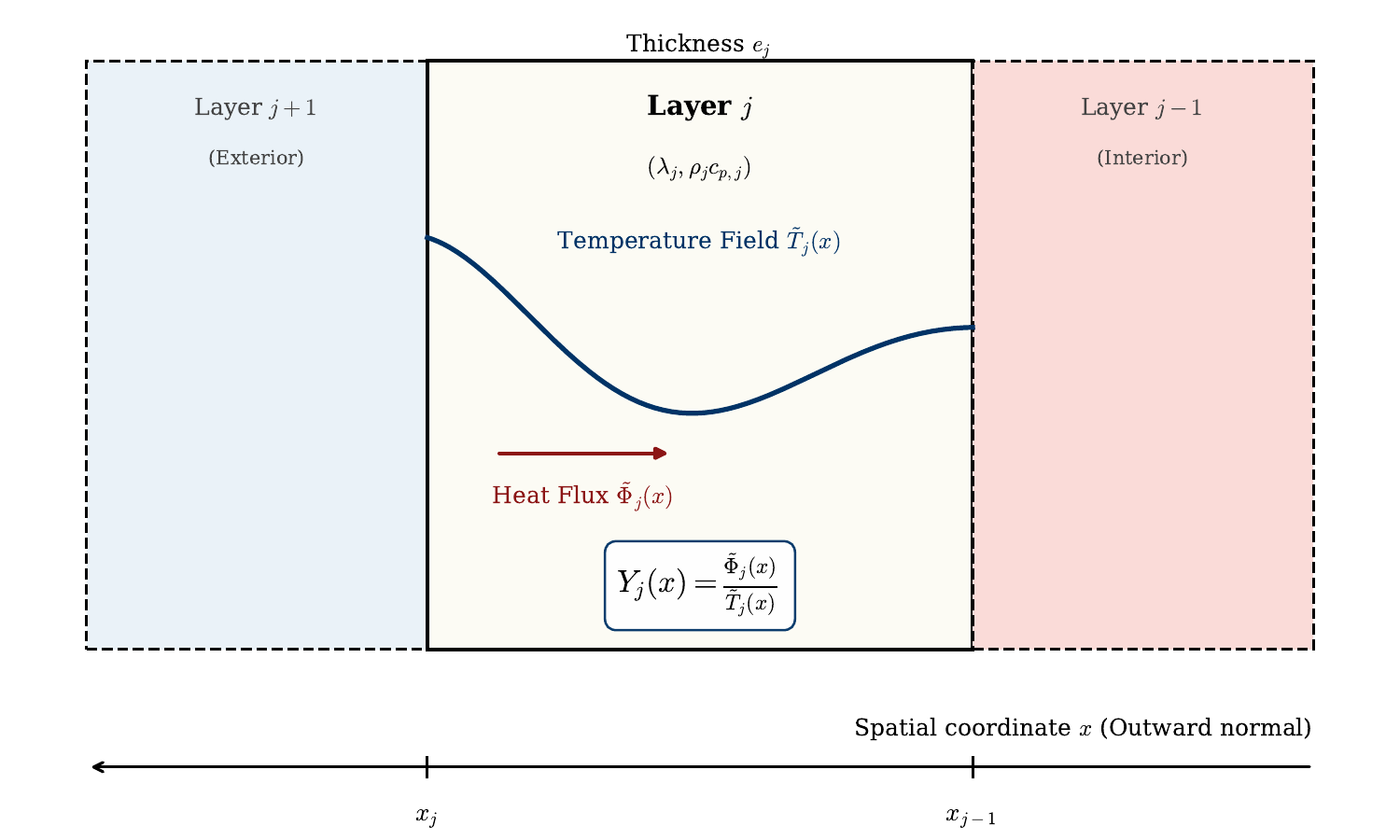}
	\caption{Conceptual schematic of the dynamic thermal admittance mapping across a homogeneous stratum $j$. The macroscopic spatial formulation (governed by the Helmholtz equation) is mathematically compressed into a local impedance ratio $Y_j(x)$. This framework maps the continuous relationship between the dissipative thermal wave $\tilde{T}_j(x)$ and the conductive heat flux $\tilde{\Phi}_j(x)$ along the inward normal coordinate $x$, bounded between the external interface $x_{j+1}$ and the internal interface $x_{j-1}$.}
	\label{fig:admittance_schematic}
\end{figure}

Standard field formulations often yield numerical instabilities when evaluating damped wave functions in thick or resistive media. An impedance-based approach avoids this evaluation. As illustrated in the conceptual schematic (Figure~\ref{fig:admittance_schematic}), the dynamic thermal admittance $Y_j(x)$ is defined as the local complex ratio of the conductive heat flux to the temperature field \cite{davies1973thermal}:
\begin{equation}
	Y_j(x) = - \frac{\tilde{\Phi}_j(x)}{\tilde{T}_j(x)} = \frac{\lambda_j}{\tilde{T}_j(x)}\frac{d\tilde{T}_j(x)}{dx}
	\label{eq:admittance_def}
\end{equation}

By differentiating Eq.~\eqref{eq:admittance_def} with respect to $x$ and substituting the second derivative from the Helmholtz equation \eqref{eq:helmholtz}, the second-order linear dynamics are mapped onto a first-order non-linear spatial Riccati equation \cite{bellman1959invariant, pagneux1996study, chiroiu2001propagation, mufti1969solution}:
\begin{equation}
	\frac{dY_j(x)}{dx} = \lambda_j q_j^2 - \frac{Y_j(x)^2}{\lambda_j}
	\label{eq:riccati}
\end{equation}
While the Riccati equation is mathematically non-linear with respect to the admittance state variable $Y_j(x)$, the underlying physical medium governed by this zero-order baseline remains strictly linear and time-invariant (constant $\lambda_j$ and $c_{p,j}$).

\subsection{Recursive Propagator}

Equation~\eqref{eq:riccati} governs the continuous spatial evolution of the admittance across layer $j$. Analytical integration over the layer thickness $e_j$ yields a Möbius transformation \cite{reid1972riccati}. In the context of the electro-thermal analogy, this recursive propagator is mathematically isomorphic to the standard impedance translation equation of a lossy electrical transmission line. By propagating the solution outward from the internal interface ($x_{j-1}$) to the external interface ($x_j$), the admittance is updated as follows:
\begin{equation}
	Y_j = Y_{c,j} \frac{Y_{j-1} + Y_{c,j} \tanh(q_j e_j)}{Y_{c,j} + Y_{j-1} \tanh(q_j e_j)}
	\label{eq:propagator}
\end{equation}
Here, $Y_{c,j}$ represents the intrinsic characteristic thermal admittance of the semi-infinite medium $j$:
\begin{equation}
	Y_{c,j} = \lambda_j q_j = \sqrt{i\omega \rho_j c_{p,j} \lambda_j}
\end{equation}

As established in wave propagation formalisms, this recursive sequence can be formally expanded into a continued fraction, establishing a strict mathematical link between discrete stratified assemblies and continuous gradient-index (GRIN) models. 

Standard transfer matrix methods (TMM) \cite{davies1973thermal, redheffer1962relation} rely on global transmission matrices involving $\sinh(q_j e_j)$ and $\cosh(q_j e_j)$. The recursive Möbius transformation in Eq.~\eqref{eq:propagator} is mathematically self-normalizing. As the dissipative argument increases ($\text{Re}(q_j e_j) \to \infty$), such as under high-frequency fluctuations or within thick insulation, the continuous analytical limit yields $\tanh(q_j e_j) \to 1$. This mathematically bounds the propagated admittance $Y_j$ to the characteristic admittance $Y_{c,j}$. However, as detailed in Section 3, the direct native floating-point evaluation of this hyperbolic function remains susceptible to catastrophic arithmetic overflow, necessitating a strict exponential reformulation to guarantee unconditional computational stability.

\subsection{System Closure and the Superposition Principle}

Evaluation of the global dynamic response requires appropriate boundary conditions. Because the governing Helmholtz equation of the unperturbed homogeneous medium is a linear operator, the global zero-order thermodynamic state of the envelope under simultaneous dual excitation (external meteorological forcing and internal setpoint variations) is resolved via the linear superposition of two independent states. The subsequent integration of physical non-linearities (moisture and radiation), which inherently violate the superposition principle, will be strictly confined to the perturbative equivalent source terms developed in Section 5. This strict mathematical decoupling preserves the $\mathcal{O}(N)$ computational scalability of linear spectral methods while evaluating non-linear boundary thermodynamics.

\subsubsection*{State A: External Meteorological Forcing}

To evaluate the transmission of the external weather through the envelope, the internal setpoint is temporarily defined as the zero-potential reference ($\tilde{T}_{in} = 0$). The internal environment is defined by a combined convective-radiative heat transfer coefficient $h_{int}$, which initializes the recursive admittance mapping at the innermost surface ($j=0$):
\begin{equation}
	Y_0 = h_{int}
	\label{eq:init_admittance}
\end{equation}

Sequential application of the Möbius propagator from the interior ($j=1$) to the exterior ($j=N$) maps the thermal admittance across the envelope, yielding the global driving-point admittance $Y_N$ at the external boundary. 

The external forcing is represented by the harmonic sol-air temperature $\tilde{T}_{sa}(\omega)$, which aggregates ambient temperature and incident solar radiation. Based on the external surface coefficient $h_{ext}$ and flux continuity, the temperature at the outermost surface $\tilde{T}_{surf}^{(A)}(\omega)$ is:
\begin{equation}
	\tilde{T}_{surf}^{(A)}(\omega) = \frac{h_{ext}}{h_{ext} + Y_N} \tilde{T}_{sa}(\omega)
\end{equation}

Determining the internal surface temperature $\tilde{T}_{si}^{(A)}(\omega)$ induced by this external forcing requires propagating the surface state backward through the strata. Based on transmission matrix relations between contiguous interfaces, spatial attenuation across layer $j$ is governed by the complex transfer factor $g_j = \tilde{T}_{j-1} / \tilde{T}_j$ \cite{athienitis1986analytical}. 

To prevent floating-point overflow during numerical evaluation, the transfer factor is reformulated to eliminate positive exponential arguments. Using the previously computed admittance $Y_{j-1}$, a bounded expression for layer attenuation is obtained:
\begin{equation}
	g_j = \frac{2 e^{-q_j e_j}}{\left( 1 + \frac{Y_{j-1}}{Y_{c,j}} \right) + e^{-2q_j e_j} \left( 1 - \frac{Y_{j-1}}{Y_{c,j}} \right)}
	\label{eq:transfer_factor}
\end{equation}

Since $\text{Re}(q_j e_j) > 0$, the exponential terms $e^{-q_j e_j}$ and $e^{-2q_j e_j}$ remain within the complex unit disk. The global thermal transfer function $G(\omega)$ is the product of the individual layer transfer factors:
\begin{equation}
	G(\omega) = \prod_{j=1}^{N} g_j
\end{equation}

The internal surface temperature response induced by the meteorological boundary is therefore:
\begin{equation}
	\tilde{T}_{si}^{(A)}(\omega) = G(\omega) \tilde{T}_{surf}^{(A)}(\omega)
	\label{eq:closure_stateA}
\end{equation}

\subsubsection*{State B: Internal Setpoint Forcing}

Model Predictive Control (MPC) strategies frequently dictate dynamic internal setpoints ($\tilde{T}_{in}(\omega) \neq 0$) to optimize load shifting. To evaluate the envelope's reaction to internal HVAC excitation, the superposition applies symmetrically. The exterior node is set to zero-potential ($\tilde{T}_{sa} = 0$). 

The spatial Riccati equation is initialized at the external interface ($\overleftarrow{Y}_N = h_{ext}$) and propagated inward to the internal surface $j=0$, yielding the backward driving-point admittance $\overleftarrow{Y}_0$. To strictly conserve the sign convention of the spatial Riccati operator (Eq.~\eqref{eq:riccati}) defined along the outward normal, this inward mapping operates under a local coordinate inversion ($\xi = e_j - x$). A conjugate transfer function $H(\omega)$, parameterized by this backward admittance field, is subsequently derived to compute the internal surface response $\tilde{T}_{si}^{(B)}(\omega)$ induced by the HVAC system.

\subsubsection*{Global Thermodynamic State}

The internal surface temperature is defined as the sum of both boundary interactions:
\begin{equation}
	\tilde{T}_{si}(\omega) = \tilde{T}_{si}^{(A)}(\omega) + \tilde{T}_{si}^{(B)}(\omega)
	\label{eq:closure_final}
\end{equation}

This closure scheme establishes a $2 \times \mathcal{O}(N)$ computational pathway, preserving the linear scalability of the framework \cite{meyer1973initial}. 

\textit{Note: For the transient benchmarks evaluated in the subsequent sections of this paper, the internal boundary is maintained at a constant setpoint (e.g., $20\,^{\circ}\mathrm{C}$). For all dynamic harmonics ($\omega > 0$), the spectral excitation $\tilde{T}_{in}(\omega) = 0$. This renders the State B dynamic contribution null, isolating the evaluation to State A to assess the non-linear physical interactions occurring at the meteorological boundary.}

	\section{Algorithmic Stability and Bounded Admittance Mapping}
	
	\subsection{Exponential Reformulation of the Admittance Mapping}
	
	While the Möbius transformation in Eq.~\eqref{eq:propagator} provides an analytical mapping of the thermal admittance, direct computational evaluation of the hyperbolic tangent function ($\tanh$) remains susceptible to internal overflow in standard numerical libraries under high dissipative arguments. To ensure absolute algorithmic stability across all spatial dimensions and frequency spectra, the propagator is algebraically reformulated to systematically avoid the evaluation of positive exponential arguments.
	
	Using the exponential definition of the hyperbolic tangent, $\tanh(z) = (1 - e^{-2z}) / (1 + e^{-2z})$, the recursive outward layer update is rewritten strictly in terms of the complex attenuation factor $e^{-2q_j e_j}$:
	\begin{equation}
		Y_j = Y_{c,j} \frac{Y_{j-1} \left( 1 + e^{-2q_j e_j} \right) + Y_{c,j} \left( 1 - e^{-2q_j e_j} \right)}{Y_{c,j} \left( 1 + e^{-2q_j e_j} \right) + Y_{j-1} \left( 1 - e^{-2q_j e_j} \right)}
		\label{eq:propagator_exponential}
	\end{equation}
	
	Since the complex thermal wave vector $q_j$ retains a strictly positive real part for any dynamic regime ($\omega > 0$), the geometric argument $-2q_j e_j$ maintains a negative real component. The magnitude of the exponential term remains bounded ($|e^{-2q_j e_j}| \le 1$). For high dissipative thermal thicknesses where $\text{Re}(q_j e_j) \gg 1$, the exponential term approaches zero, circumventing standard 64-bit floating-point limits. The propagated admittance $Y_j$ natively converges to the characteristic admittance $Y_{c,j}$, mathematically recovering the behavior of a semi-infinite medium. 
	
	Physically, this bounded behavior analytically replicates the thermal penetration depth limit. When the physical layer thickness drastically exceeds the active penetration depth of the thermal wave ($e_j \gg \sqrt{2\alpha_j/\omega}$), the exponential formulation converges to the semi-infinite boundary condition without requiring manual thresholding or conditional algorithmic branching.
	
	\subsection{Numerical Limitations of the Transfer Matrix Method}
	
	In applied building physics, the one-dimensional propagation of thermal waves is conventionally resolved using the standard Transfer Matrix Method (TMM) \cite{davies1973thermal}. The formulation relies on backward-mapping the thermal state vector across a homogeneous layer $j$ from the outer interface $x_j$ to the inner interface $x_{j-1}$ via a transmission matrix $M_j$:
	$$
	\begin{bmatrix} T(x_{j-1}) \\ \Phi(x_{j-1}) \end{bmatrix} = \begin{bmatrix} \cosh(q_j e_j) & \frac{1}{\lambda_j q_j} \sinh(q_j e_j) \\ \lambda_j q_j \sinh(q_j e_j) & \cosh(q_j e_j) \end{bmatrix} \begin{bmatrix} T(x_j) \\ \Phi(x_j) \end{bmatrix}
	$$
	
	As the dimensionless thermal thickness $\text{Re}(q_j e_j)$ increases (typically in highly resistive envelopes or under high-frequency transient forcing), the hyperbolic functions converge asymptotically toward exponentials ($\approx \frac{1}{2}\exp(q_j e_j)$). While analytically exact, this standard formulation presents two structural limits when evaluated using IEEE 754 64-bit floating-point arithmetic \cite{goldberg1991every, higham2002accuracy}. 
	
	First, the theoretical determinant of the transmission matrix is strictly unity. Numerically, evaluating this determinant for large arguments requires subtracting two nearly identical floating-point numbers ($\frac{1}{4}e^{2q_j e_j} - \frac{1}{4}e^{2q_j e_j}$). This operation exhausts the finite mantissa of the 64-bit precision ($\epsilon \approx 2.22 \times 10^{-16}$), causing a catastrophic loss of linear independence and rendering the global system matrix ill-conditioned \cite{higham2002accuracy}.
	
	Second, when the thermal thickness reaches $\text{Re}(q_j e_j) \ge 709$, the exponential term exceeds the maximum representable finite number in 64-bit precision ($\approx 1.79 \times 10^{308}$). At this threshold, the evaluation triggers an arithmetic overflow, returning non-finite values (NaN) and halting the resolution \cite{goldberg1991every}. While advanced wave propagation fields often circumvent this divergence by substituting the TMM with the Scattering Matrix Method (SMM), this intervention demands heavy block-matrix operations (Redheffer star products) to map incoming and outgoing states. The unmodified TMM, widely utilized for its sequential simplicity in applied building physics, remains inherently unstable under these dissipative regimes.
	
\begin{figure}[htbp]
		\centering
		\includegraphics[width=0.95\textwidth]{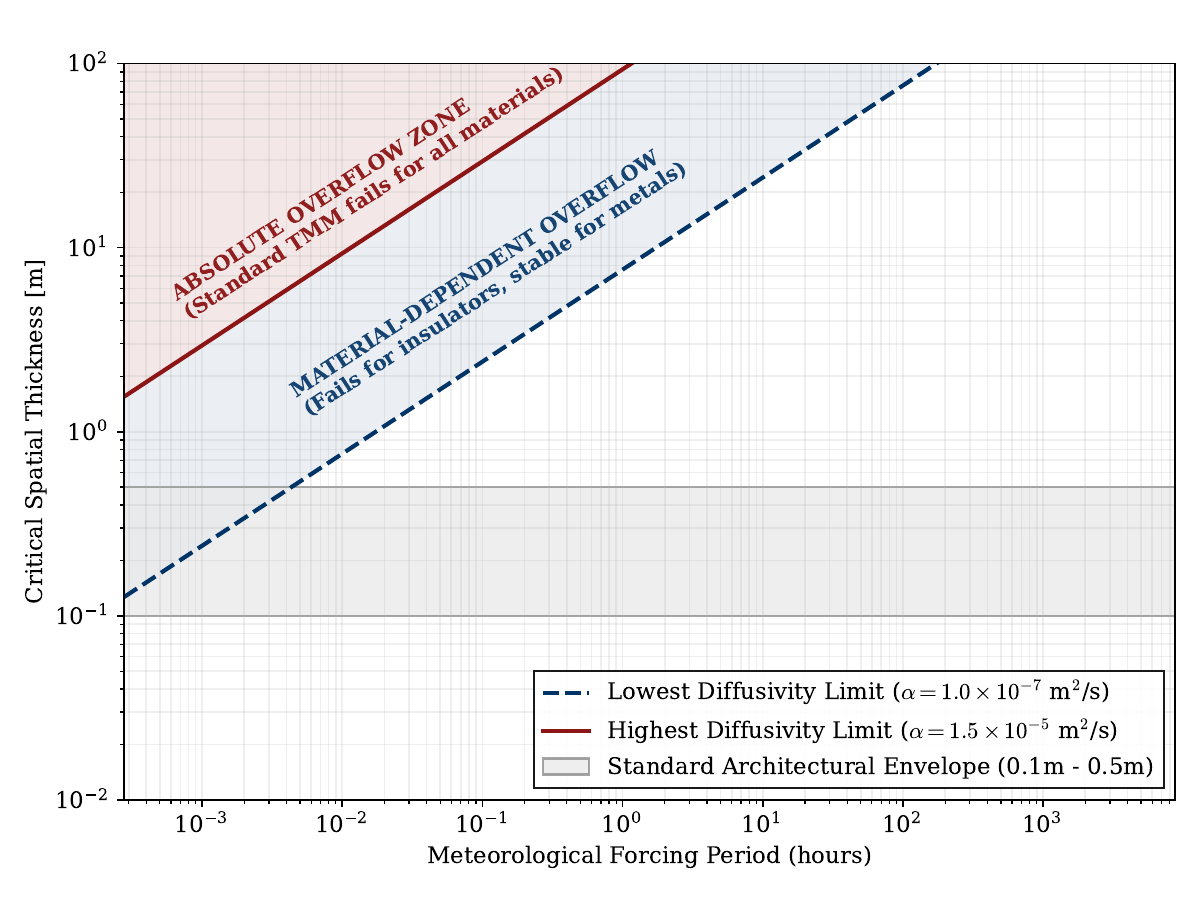}
		\caption{Analytical phase-space defining the 64-bit floating-point stability limits of the standard Transfer Matrix Method (TMM). The boundaries encompass the complete spectrum of building materials, from the lowest diffusivity limit (insulators, $\alpha = 1.0 \times 10^{-7}$ m$^2$/s) to the highest diffusivity limit (structural metals, $\alpha = 1.5 \times 10^{-5}$ m$^2$/s). The shaded areas identify the spatial dimensions and forcing periods where standard matrix evaluation triggers exponential overflow ($\text{Re}(q_j e_j) \ge 709$), returning non-finite values. The proposed bounded admittance mapping ($|e^{-2q_j e_j}| \le 1$) remains completely immune to this divergence across the entire plotted domain.}
		\label{fig:tmm_phase_space}
	\end{figure}

	Figure~\ref{fig:tmm_phase_space} delineates this critical $e^{709}$ overflow boundary across the meteorological frequency spectrum for the standard range of building materials. While the TMM computes low-frequency variations (periods $> 100$ hours) without issue, it systematically overflows within standard architectural envelope dimensions (0.1 m to 0.5 m) when evaluating high-frequency transients or stochastic sensor noise (periods approaching $1$ to $10$ seconds), rendering it ill-suited for fast dynamic MPC boundary fluctuations. The bounded Riccati reformulation (Eq.~\eqref{eq:propagator_exponential}) remains structurally insulated from this phase-space divergence.
	
	\subsection{Benchmarking Deep Ground Coupling Dynamics}
	
\begin{figure}[htbp]
		\centering
		\includegraphics[width=\textwidth]{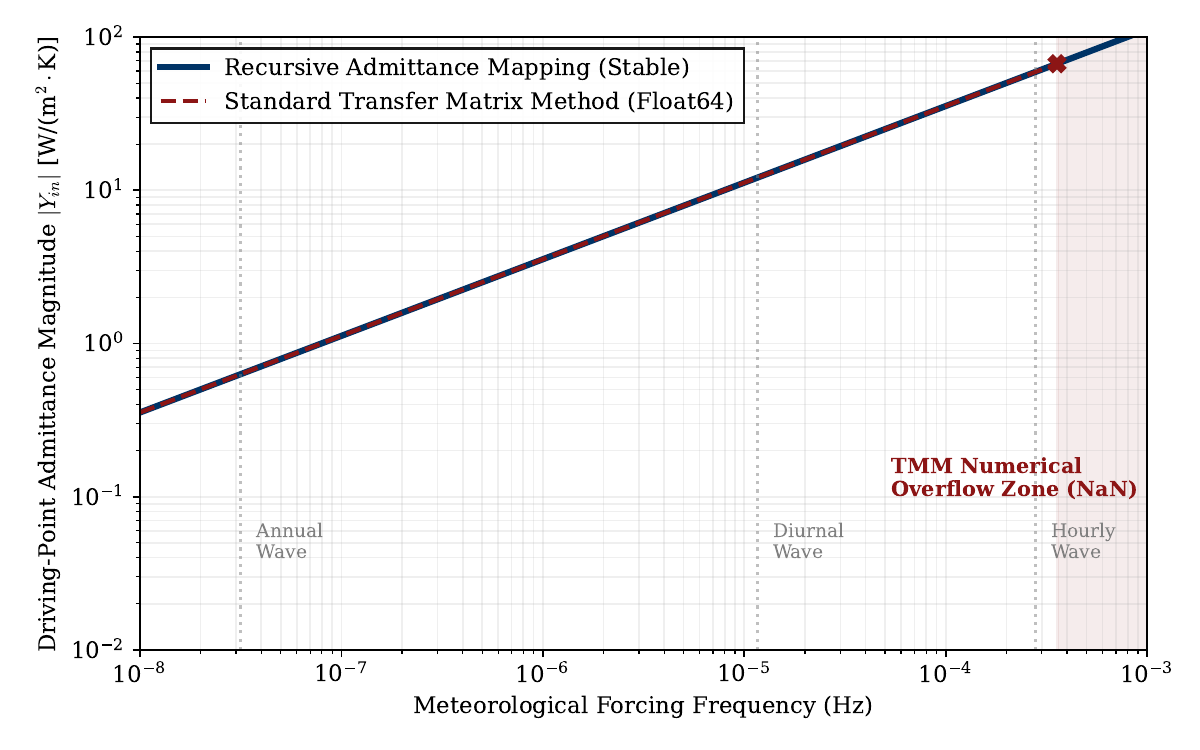}
		\caption{Dynamic thermal admittance magnitude $|Y_0|$ evaluated at the interior surface of a slab-on-grade foundation coupled with a 15-meter deep geothermal inertia layer. In the sub-hourly high-frequency regime ($f \ge 10^{-3}$ Hz), the standard Transfer Matrix Method (dashed red) evaluates positive diverging exponentials (e.g., $e^{+850}$) and crashes due to 64-bit overflow. The recursive Riccati mapping (solid blue) evaluates strictly bounded negative exponentials ($e^{-850} \to 0$), smoothly recovering the semi-infinite medium asymptote without numerical instability.}
		\label{fig:ground_coupling_breakdown}
	\end{figure}

	The numerical stability of the proposed formulation is evaluated in a deep ground coupling scenario, which represents a highly dissipative boundary condition. Figure~\ref{fig:ground_coupling_breakdown} compares the frequency response of a 15-meter deep geothermal mass computed using the recursive admittance framework and the standard TMM formulation.
	
	In the low-frequency limit ($f < 10^{-5}$ Hz, corresponding to seasonal and monthly thermal waves), spatial attenuation is low, and the TMM matrices evaluate without overflow. The spectral superposition of both models in this region confirms that the recursive admittance mapping introduces no mathematical approximations and strictly replicates the theoretical thermodynamic baseline.
	
	As the forcing frequency increases toward sub-hourly fluctuations ($f \ge 10^{-3}$ Hz) typically required for high-resolution Model Predictive Control, the active thermal penetration depth shrinks to a few centimeters. Automated optimization solvers routinely evaluate extreme or unfiltered architectural configurations without manual physical checks. Under these high-frequency conditions, standard TMM solvers process diverging positive exponential terms (e.g., $e^{+850}$) within the transmission matrices, violently exceeding the 64-bit IEEE 754 limit and terminating the systemic computation. 
	
	The exponential Riccati reformulation evaluates the identical physical scenario by smoothly computing $e^{-850} \to 0$. This evaluation inherently collapses the system to a valid semi-infinite boundary. While resolving sub-hourly fluctuations at a depth of 15 meters is  thermodynamically irrelevant as the thermal wave is strictly annihilated within the first few centimeters, this scenario constitutes a rigorous numerical stress-test. It demonstrates that the recursive Riccati framework maintains mathematical integrity where standard matrix-based methods fail. This structural stability guarantees that automated MPC solvers, which blindly explore parameter phase-spaces during optimization, remain perfectly immune to unphysical architectural configurations or high-frequency sensor noise.
	
	\section{Algorithmic Completeness and Time-Domain Reconstruction}
	
	\subsection{Spectral Discretization and the Stationary Limit}
	
	The mathematical developments presented in the previous sections operate in the harmonic domain. Building energy management systems, such as Model Predictive Control (MPC), require transient time-domain evaluations of stochastic weather profiles. Integrating the strictly $\mathcal{O}(N)$ spatial frequency-domain propagator into a time-domain simulation requires a spectral reconstruction methodology \cite{athienitis1986analytical}.
	
	A discrete time-domain meteorological forcing, represented by the sol-air temperature $T_{sa}(t)$, is sampled at an interval $\Delta t$ over a finite predictive horizon $\tau$. This signal is mapped into the frequency domain using the Fast Fourier Transform (FFT) algorithm, operating in $\mathcal{O}(M \log M)$ temporal complexity, where $M = \tau / \Delta t$ is the total number of temporal nodes. Let $\omega_k = 2\pi k / \tau$ be the discrete angular frequencies. To rigorously preserve the absolute thermodynamic dimension (Kelvins) of the spectral amplitudes and strictly isolate them from algorithmic energy inflation, the discrete operator is inherently normalized:
	\begin{equation}
		\tilde{T}_{sa}(\omega_k) = \frac{1}{M} \sum_{n=0}^{M-1} T_{sa}(n \Delta t) e^{-i \omega_k n \Delta t}
	\end{equation}
	
	Because the input signal $T_{sa}(t)$ is real-valued, its Fourier spectrum exhibits Hermitian symmetry ($\tilde{T}_{sa}(-\omega) = \tilde{T}_{sa}^*(\omega)$). Furthermore, the complex thermal wave vector inherently satisfies $q(-\omega) = q^*(\omega)$, guaranteeing that the resulting dynamic admittance and global transfer functions strictly preserve this Hermitian conjugation. Consequently, the $\mathcal{O}(N)$ recursive admittance mapping is evaluated only for the positive half of the frequency spectrum ($k \in [0, M/2]$), yielding a global spatio-temporal system complexity of $\mathcal{O}(M \log M + N \cdot M)$ while mathematically ensuring a strictly real time-domain reconstruction.
	
\begin{figure}[htbp]
		\centering
		\includegraphics[width=0.95\textwidth]{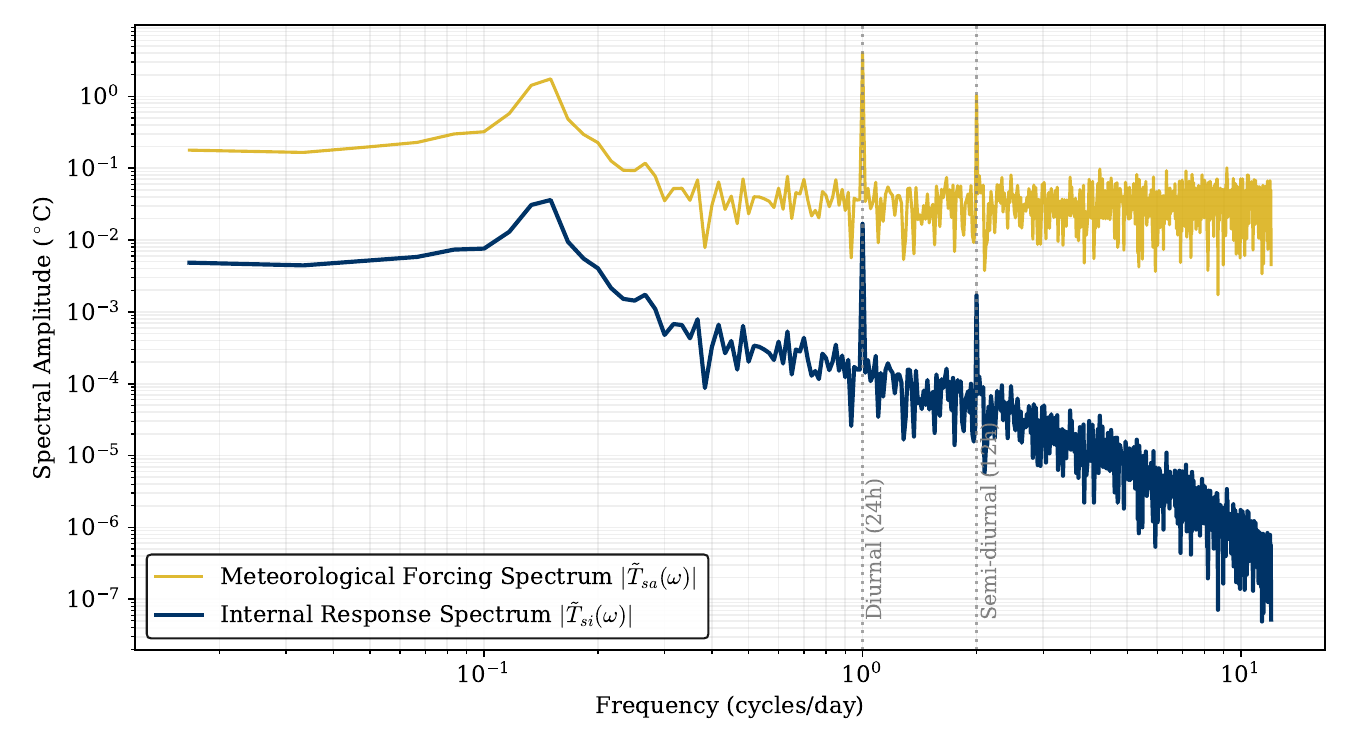}
		\caption{Frequency-domain amplitude spectrum of a stochastic meteorological forcing and its corresponding internal thermal response across a heavy composite envelope. The macroscopic thermal diffusion acts as a low-pass filter, exponentially attenuating high-frequency noise and diurnal harmonics ($\omega > 0$). The fundamental static component ($\omega = 0$) is resolved via the Laplace limit to ensure energy conservation.}
		\label{fig:spectral_filter}
	\end{figure}

	As illustrated in Figure~\ref{fig:spectral_filter}, evaluating this spectrum highlights the low-pass filtering nature of the building envelope. The macroscopic thermal diffusion operator exponentially attenuates high-frequency meteorological noise and rapid diurnal fluctuations. Because dynamic harmonics ($\omega_k > 0$) are heavily damped, they are processed using the bounded exponential propagator (Eq.~\eqref{eq:propagator_exponential}). The unattenuated stationary component ($k=0 \implies \omega_0 = 0$) dictates the base energy load and requires a specific limit analysis to ensure algorithmic completeness.
	
	In this static limit ($\omega \to 0$), the complex thermal wave vector vanishes ($q_j \to 0$). The spatial Riccati equation (Eq.~\eqref{eq:riccati}) rigorously collapses to an elementary separable differential equation:
	\begin{equation}
		\frac{dY_j(x)}{dx} = - \frac{Y_j(x)^2}{\lambda_j}
	\end{equation}
	
	Separating the variables and integrating directly across the spatial domain of the stratum $[x_{j-1}, x_j]$ yields the steady-state thermal resistance update rule:
	\begin{align}
		\int_{Y_{j-1}}^{Y_j} \frac{dY}{Y^2} &= -\int_{x_{j-1}}^{x_j} \frac{dx}{\lambda_j} \\
		-\left[ \frac{1}{Y_j} - \frac{1}{Y_{j-1}} \right] &= -\frac{e_j}{\lambda_j} \\
		Y_j &= \frac{1}{\frac{1}{Y_{j-1}} + \frac{e_j}{\lambda_j}}
		\label{eq:propagator_stationary_base}
	\end{align}
	
	This direct analytical integration demonstrates that the recursive spatial Riccati approach natively and continuously converges to steady-state macroscopic heat conduction principles without mathematical singularities or requiring arbitrary local coordinate detours.
	
	\subsection{Temporal Aliasing and Transient Error Bounding}
	
	The transition from the frequency domain to the time domain via the Inverse Discrete Fourier Transform (iDFT) assumes that the meteorological forcing signal $T_{sa}(t)$ is periodic. This boundary condition results in a circular convolution between the forcing signal and the thermal system \cite{clarke2001energy}. If the initial thermal state of the envelope differs from its final state at the end of the simulation window $\tau$, a temporal wrap-around artifact occurs, introducing errors in the predicted internal temperature response at the beginning of the simulation. 
	
	To mitigate this temporal aliasing, the dissipative nature of the thermal diffusion operator is utilized. The transient thermal impulse response $h(t)$ of a multi-layered envelope decays exponentially according to the dominant macroscopic thermal relaxation time constant of the wall, $\tau_{dom}$.
	
	The residual truncation error $\varepsilon(t)$ caused by the circular convolution is bounded by this exponential decay. By prepending a meteorological sequence of duration $t_{wu}$ (a warm-up period) prior to the active predictive horizon, the aliasing artifact at the beginning of the predictive window ($t = t_{wu}$) is attenuated \cite{clarke2001energy}:
	\begin{equation}
		\varepsilon(t_{wu}) \le \mathcal{O}\left( \exp\left(-\frac{t_{wu}}{\tau_{dom}}\right) \right)
	\end{equation}
	
	Setting a warm-up period of $t_{wu} \ge 3\tau_{dom}$ bounds the initial state artifact to less than $5\%$ ($e^{-3}$), and $t_{wu} \ge 5\tau_{dom}$ limits the residual error to below $1\%$ ($e^{-5}$). 
	
	To empirically quantify this truncation error, a transient benchmark is evaluated on a highly inertial envelope (40 cm of solid concrete). To maximize the wrap-around artifact, a non-stationary meteorological boundary condition is constructed. The synthetic forcing signal consists of a stationary historical baseline (mean temperature of 15°C) followed by a 7-day predictive horizon featuring a severe meteorological front (a linear 10°C temperature drop). This protocol enforces a thermal discontinuity between the initial and final states of the discrete Fourier window.
	
	To prevent spectral leakage and the artificial high-frequency harmonics (Gibbs phenomenon) generated by this temporal discontinuity, a standard linear detrending is applied to the input signal prior to the discrete Fourier transform. The linear trend is subsequently analytically superimposed onto the final time-domain reconstruction.
	
\begin{figure}[htbp]
		\centering
		\includegraphics[width=0.85\textwidth]{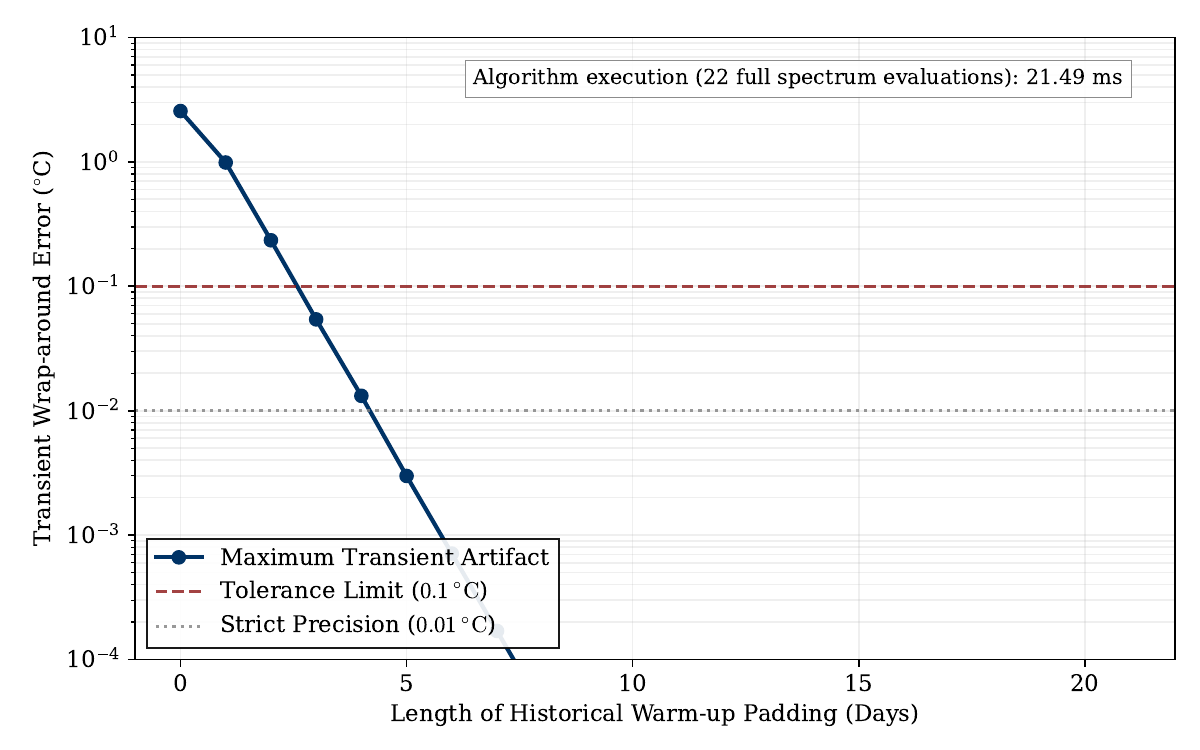}
		\caption{Exponential decay of the transient wrap-around error (temporal aliasing) as a function of the historical warm-up padding duration. The benchmark evaluates a highly inertial 40~cm massive concrete envelope subjected to a non-stationary meteorological front (a 10°C temperature drop over the predictive horizon). As dictated by macroscopic thermal diffusion, the truncation error decays exponentially, crossing the 0.01°C precision threshold after 4 days of temporal padding. The global $\mathcal{O}(M \log M + N \cdot M)$ algorithmic evaluation over the extended spectrum executes in under 25 milliseconds, demonstrating that transient bounding introduces minimal computational overhead.}
		\label{fig:spectral_aliasing}
	\end{figure}

	As illustrated in Figure~\ref{fig:spectral_aliasing}, the maximum transient error observed during the first day of the predictive window is plotted as a function of the warm-up duration $t_{wu}$. The aliasing artifact undergoes an exponential decay, driven by the macroscopic thermal relaxation time of the concrete. For this specific configuration, a temporal padding of 4 days is sufficient to suppress the transient error below a 0.01°C precision threshold. 
	
	Because the recursive admittance framework evaluates the spectrum analytically, processing this extended historical spectrum only impacts the global $\mathcal{O}(M \log M + N \cdot M)$ complexity. This continuous spatial evaluation natively bypasses the computational bottleneck of discrete methods, executing the padded transient horizon in under 25 milliseconds. This spectral padding strategy guarantees a transient response robust for MPC operations without reverting to iterative time-stepping algorithms.
	
	While this spectral padding strategy guarantees a robust transient response independent of arbitrary initial states, it inherently relies on the assumption that the envelope is in dynamic equilibrium with the prescribed historical weather. For MPC architectures requiring the direct assimilation of measured internal spatial temperature gradients at $t=0$ (e.g., via extended Kalman filters), time-domain state-space models remain structurally more adequate than frequency-domain propagators.
	
	\subsection{State-Space Dimension and Spatial Discretization Limits}
	
	To evaluate the numerical stability conditions of the discrete Finite Difference Method (FDM), the one-dimensional heat equation is analyzed: $\frac{\partial T}{\partial t} = \alpha \frac{\partial^2 T}{\partial x^2}$.
	
	Discretizing this equation using a Forward Time Centered Space (FTCS) explicit scheme yields an algebraic nodal update governed by the dimensionless grid Fourier number, $Fo = \alpha \Delta t / \Delta x^2$. To prevent the amplification of numerical errors, the Courant-Friedrichs-Lewy (CFL) stability criterion strictly enforces $Fo \leq 0.5$ \cite{courant1967partial}.
	
	For highly resistive insulation layers requiring fine spatial resolution ($\Delta x \to 0$ to prevent numerical diffusion), the CFL criterion restricts the temporal step to unmanageable sub-minute intervals, causing explicit solvers to diverge at standard MPC sampling rates (e.g., $\Delta t = 300$ s). To bypass this temporal instability, standard simulation engines exclusively deploy implicit finite difference schemes (e.g., Crank-Nicolson or Backward Euler), which are unconditionally stable. 
	
	While implicit tridiagonal matrices can be inverted efficiently in linear time using the Thomas algorithm, the structural requirement for discrete spatial nodes severely inflates the dimension of the state-space vector. Ensuring thermodynamic accuracy demands a refined spatial grid (e.g., $M_s \ge 50$ nodes per envelope). The fundamental limitation of discrete methods is therefore not temporal stability, but the systemic dimension inflation ($\mathcal{O}(\sum M_s)$) inherently required to suppress spatial truncation errors, paralyzing urban-scale stochastic optimization.
	
	Since spatial derivatives are pre-integrated analytically over the continuous domain $x \in [0, e_j]$ in the admittance mapping method, the internal spatial discretization grid ($\Delta x$) is eliminated. The spatial algorithmic complexity scales strictly as $\mathcal{O}(N)$, where $N$ is the number of macroscopic architectural strata. Because $N$ remains bounded regardless of the material's thermal resistance (e.g., $N=5$ physical layers versus $M_s > 100$ equivalent numerical nodes for thick insulation), the computational framework evaluates spatial dynamics without staircase approximation errors, remaining independent of the thermal penetration depth or the temporal resolution \cite{ouyang1991procedure}.
	
\begin{figure}[htbp]
		\centering
		\includegraphics[width=0.85\textwidth]{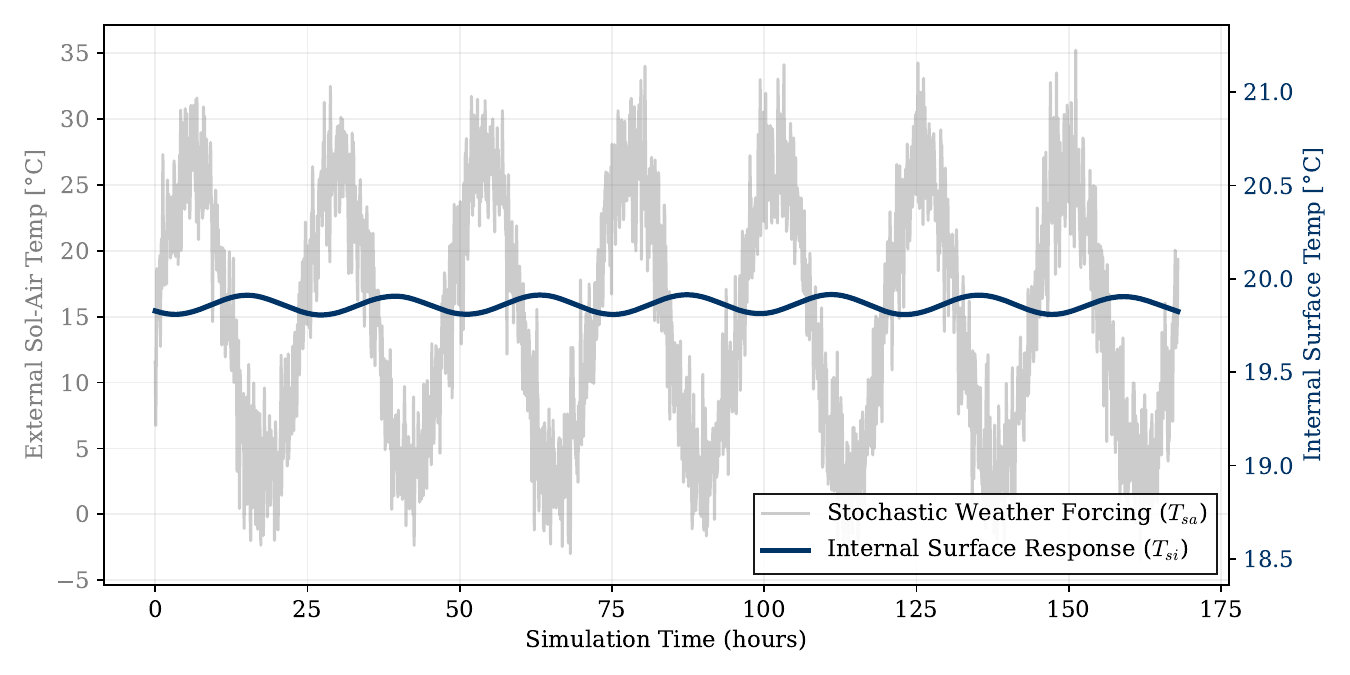}
		\caption{Transient time-domain reconstruction of a heavily insulated composite wall internal surface temperature under stochastic meteorological forcing ($\Delta t = 300$ s). The recursive admittance solver provides a stable evaluation capturing macroscopic thermal inertia without spatial discretization or implicit state-space inflation. The multi-week transient evaluation executes in under 2 ms.}
		\label{fig:transient_benchmark}
	\end{figure}

	Figure~\ref{fig:transient_benchmark} illustrates this continuous spatial evaluation for a composite wall (20 cm concrete, 15 cm EPS) under stochastic forcing at $\Delta t = 300$ s. The time-domain reconstruction achieved by the $\mathcal{O}(M \log M + N \cdot M)$ recursive admittance solver provides a stable evaluation capturing macroscopic thermal inertia without spatial discretization or implicit state-space inflation. The high-frequency noise is filtered by the exponential propagator, yielding an attenuated and phase-shifted internal surface temperature response, with the multi-week horizon executing in under 2 ms on standard computational hardware.
	
	\section{Perturbative Extension: Real-World Building Physics}
	
	\subsection{The Limits of the Homogeneous Assumption: Coupled Heat and Moisture}
	
	The baseline analytical propagator assumes material homogeneity, defined by constant thermophysical properties ($\lambda_j$, $c_{p,j}$). In applied building physics, these properties are state-dependent. While temperature-induced variations in dry insulators generally remain marginal, the presence of interstitial moisture in porous media alters the macroscopic thermodynamic response.
	
	For highly porous structural envelopes, such as Autoclaved Aerated Concrete (AAC) or wood fiberboards, the accumulation of liquid water modifies both thermal conductivity and volumetric heat capacity. According to standard hygrothermal models (e.g., ISO 10456, \cite{iso10456}), the effective conductivity scales exponentially with the volumetric moisture content $w(x)$, while the sensible heat capacity increases linearly due to the high density and specific heat of occluded liquid water. Under winter conditions, interstitial condensation often generates a continuous spatial moisture gradient across the structural layer.
	
	Modeling full dynamic coupled Heat, Air, and Moisture (HAM) transfer, including latent heat from phase changes (evaporation, freezing), requires solving complex non-linear partial differential equations. Iterative spatial resolution of these fields is computationally intensive for real-time Model Predictive Control (MPC) algorithms. To maintain algorithmic efficiency, the present scope is strictly limited to sensible moisture-dependent thermophysical degradation. A physical separation of time scales applies: the macroscopic relaxation time of capillary moisture transport spans several weeks to months, whereas thermal diffusion operates on an hourly to daily basis.
	
	Over a standard 7-day MPC predictive horizon, the spatial moisture distribution $w(x)$ can be modeled as a quasi-static internal parameter field.
	
\begin{figure}[htbp]
		\centering
		\includegraphics[width=0.85\textwidth]{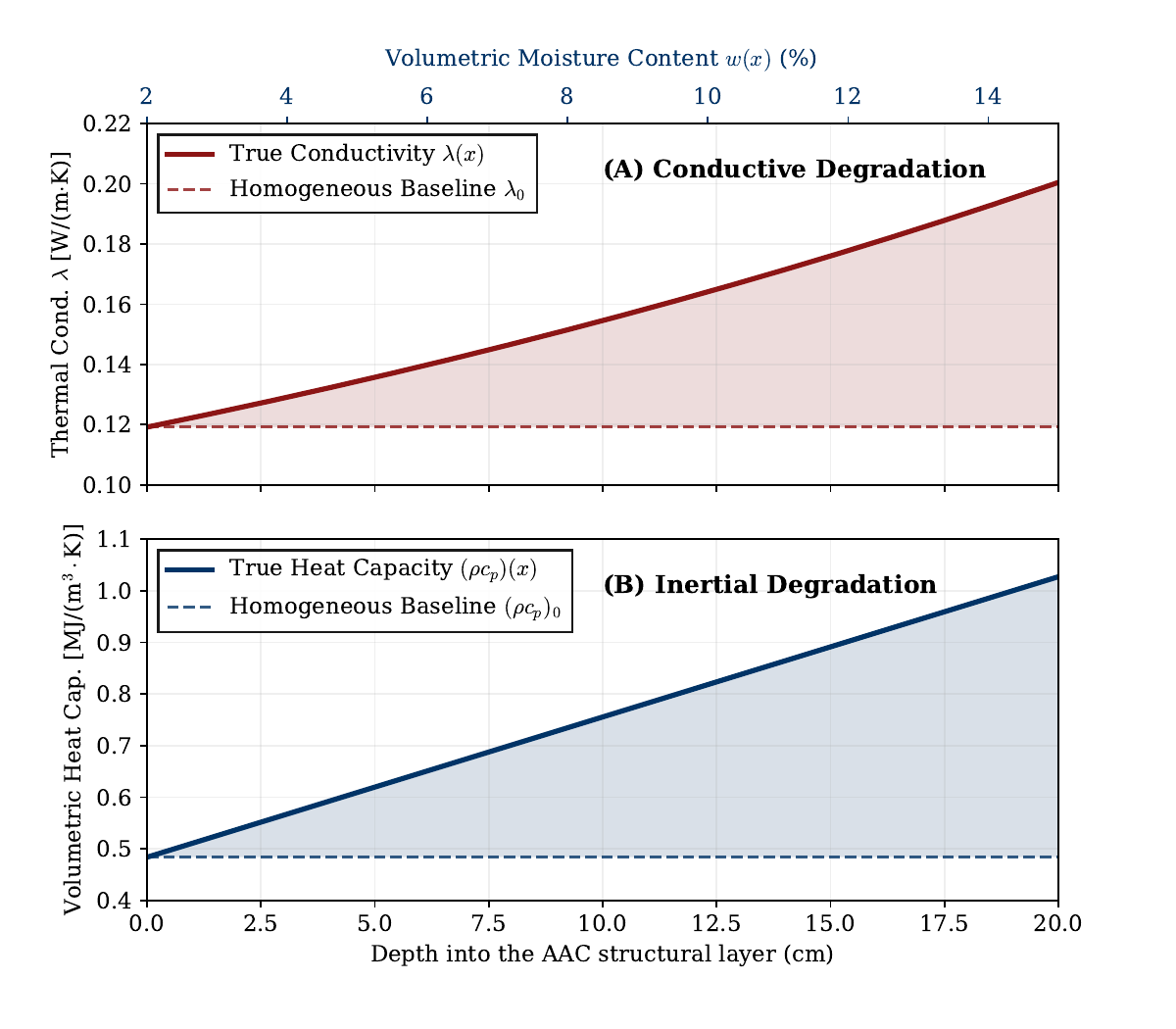}
		\caption{Spatial degradation of thermophysical properties across a 20~cm Autoclaved Aerated Concrete (AAC) structural layer subjected to a static winter moisture gradient (ranging from 2\% interior to 15\% exterior volumetric moisture content). \textbf{(A)} Thermal conductivity $\lambda(x)$ scales exponentially with moisture, diverging from the homogeneous dry baseline $\lambda_0$. \textbf{(B)} Volumetric heat capacity $(\rho c_p)(x)$ increases linearly due to the presence of occluded liquid water, deviating from the nominal baseline $(\rho c_p)_0$. The proposed perturbative framework maps these continuous spatially distributed discrepancies into the dynamic admittance computation without requiring internal structural node discretization.}
		\label{fig:property_perturbation}
	\end{figure}

	To illustrate this physical deviation, Figure~\ref{fig:property_perturbation} evaluates the spatial profile of thermophysical properties across a 20~cm AAC layer subjected to a static winter moisture accumulation. Driven by the occluded water, the true conductivity $\lambda(x)$ and volumetric heat capacity $(\rho c_p)(x)$ exhibit a continuous degradation from their nominal dry baselines ($\lambda_0$ and $(\rho c_p)_0$). As shown in Panel A, the thermal conductivity rises exponentially from $0.12$ to $0.20~\mathrm{W/(m\cdot K)}$ at the wetted boundary. Panel B illustrates that the volumetric heat capacity more than doubles, increasing linearly from $0.49$ to $1.03~\mathrm{MJ/(m^3\cdot K)}$. Standard numerical solvers require spatial discretization of the medium into sub-nodes to capture this continuous dual degradation. To bypass this computational step, the spatial Riccati framework is extended using regular perturbation theory.
	
	To formally derive the perturbed propagator, the continuous spatial Riccati equation (Eq.~\eqref{eq:riccati}) is expressed explicitly in terms of the volumetric heat capacity, using the identity $\lambda(x) q(x)^2 = i\omega (\rho c_p)(x)$:
	\begin{equation}
		\frac{dY(x)}{dx} = i\omega (\rho c_p)(x) - \frac{Y(x)^2}{\lambda(x)}
	\end{equation}
	
	A finite spatial perturbation of the thermophysical properties is defined around the nominal dry baselines:
	\begin{align}
		\lambda(x) &= \lambda_0 + \delta\lambda(x) \\
		(\rho c_p)(x) &= (\rho c_p)_0 + \delta(\rho c_p)(x) \\
		Y(x) &= Y^{(0)}(x) + Y^{(1)}(x)
	\end{align}
	where the spatially distributed discrepancies $\delta\lambda(x)$ and $\delta(\rho c_p)(x)$ represent the static moisture-driven degradation, and $Y^{(1)}(x)$ is the first-order admittance correction.
	
	Substituting these expansions into the Riccati equation yields:
	\begin{equation}
		\frac{d}{dx} \left[ Y^{(0)}(x) + Y^{(1)}(x) \right] = i\omega \left[ (\rho c_p)_0 + \delta(\rho c_p)(x) \right] - \frac{\left( Y^{(0)}(x) + Y^{(1)}(x) \right)^2}{\lambda_0 + \delta\lambda(x)}
	\end{equation}
	
	Applying a first-order Taylor expansion to the non-linear fractional term, $(1 + \delta\lambda/\lambda_0)^{-1} \approx 1 - \delta\lambda/\lambda_0$, and neglecting the second-order terms ($Y^{(1)} \delta\lambda$ and $(Y^{(1)})^2$), the right-hand side is linearized. Isolating the zero-order terms recovers the nominal Riccati equation for the dry homogeneous medium. The remaining first-order terms form a linear ordinary differential equation governing the admittance correction $Y^{(1)}(x)$:
	\begin{equation}
		\frac{dY^{(1)}(x)}{dx} + \frac{2 Y^{(0)}(x)}{\lambda_0} Y^{(1)}(x) = \mathcal{S}(x)
		\label{eq:riccati_perturbed}
	\end{equation}
	
	The equivalent harmonic source term $\mathcal{S}(x)$ isolates the physical non-linearities, evaluated entirely using the known zero-order fields:
	\begin{equation}
		\mathcal{S}(x) = \frac{\delta \lambda(x)}{\lambda_0^2} \left[ Y^{(0)}(x) \right]^2 + i\omega \delta(\rho c_p)(x)
	\end{equation}
	
	From a thermodynamic perspective, this source term mathematically decouples the two fundamental mechanisms of heat transfer altered by the presence of liquid water. The first term, proportional to $+\delta\lambda$, represents the perturbation of the thermal resistance. It dictates the amplitude modulation of the thermal wave as it propagates through a medium that has become highly conductive. The second term, $+i\omega \delta(\rho c_p)$, represents the perturbation of the thermal capacity. Its strictly imaginary nature ($i\omega$) indicates that the sensible heat storage of the occluded water induces a temporal phase shift, delaying the propagation of the wave. By integrating both terms, the formulation captures both the accelerated heat transfer and the increased thermal inertia without requiring internal structural node discretization.
	
	Regular perturbation theory formally requires the local fractional perturbation to remain strictly bounded ($\delta\lambda / \lambda_0 \ll 1$) to justify the first-order Taylor expansion $(1 + \delta\lambda/\lambda_0)^{-1} \approx 1 - \delta\lambda/\lambda_0$. In the evaluated hygrothermal scenario, the fractional discrepancy reaches a severe local magnitude of $+0.66$ at the wetted exterior boundary. 
	
	To formally quantify the validity of this expansion under such extreme local gradients, the exact algebraic residual of the Taylor truncation is derived. The first-order admittance correction is governed by a linear ordinary differential equation (Eq.~\eqref{eq:riccati_perturbed}). Substituting the fundamental definition of the zero-order admittance, $Y^{(0)}(x) = \frac{\lambda_0}{T^{(0)}(x)} \frac{dT^{(0)}(x)}{dx}$, the integrating factor of this system is analytically identified as the square of the unperturbed temperature field, $\left[T^{(0)}(x)\right]^2$. Multiplying the exact local algebraic error by this integrating factor, the global truncation error $\mathcal{E}_{trunc}$ introduced into the macroscopic propagator is strictly defined by the definite integral:
	\begin{equation}
		\mathcal{E}_{trunc} = \int_{0}^{e_j} \left[ T^{(0)}(z) Y^{(0)}(z) \right]^2 \frac{1}{\lambda_0} \left( \frac{\epsilon(z)^2}{1 + \epsilon(z)} \right) dz
		\label{eq:error_bound}
	\end{equation}
	where $\epsilon(z) = \delta\lambda(z)/\lambda_0 = e^{\beta z} - 1$. While the local algebraic error reaches $\approx 26\%$ at the strict exterior boundary under extreme moisture accumulation, the systemic robustness of the framework is mathematically guaranteed by the integral evaluation. Because the true spatial perturbation $\epsilon(z)$ strictly follows an exponential decay toward the interior, the quadratic residual $\epsilon(z)^2$ drastically compresses the error weight across the structural domain. The localized geometric divergence at the boundary is analytically absorbed by the macroscopic integration, maintaining the global transmission error at an asymptotic minimum ($\approx 0.44\%$, as demonstrated in the subsequent computational benchmark) without necessitating intractable higher-order ($Y^{(2)}$) expansions.

	\subsection{Algebraic Resolution and Continuous Integration}
	
	Applying the previously identified integrating factor, $\left[T^{(0)}(x)\right]^2$, to Eq.~\eqref{eq:riccati_perturbed} transforms the left-hand side into an exact differential:
	\begin{equation}
		\frac{d}{dx} \left[ \left( T^{(0)}(x) \right)^2 Y^{(1)}(x) \right] = \left( T^{(0)}(x) \right)^2 \mathcal{S}(x)
	\end{equation}
	
	Integrating this differential across the $j$-th layer defines the recursive propagator for the perturbative correction:
	\begin{equation}
		Y^{(1)}(x_j) = \left( \frac{T^{(0)}(x_{j-1})}{T^{(0)}(x_j)} \right)^2 Y^{(1)}(x_{j-1}) + \frac{1}{\left(T^{(0)}(x_j)\right)^2} \int_{0}^{e_j} \left( T^{(0)}(z) \right)^2 \mathcal{S}(z) dz
		\label{eq:propagator_perturbative}
	\end{equation}
	
	To implement the perturbative correction without resorting to spatial discretization, the continuous integrand is mapped from the global coordinate $x \in [x_{j-1}, x_j]$ to a local stratum coordinate $z = x - x_{j-1}$ where $z \in [0, e_j]$. By using the fundamental definition of the thermal admittance, the algebraic manipulation cancels the temperature denominator, significantly simplifying the continuous integrand $I(z) = (T^{(0)}(z))^2 \mathcal{S}(z)$:
	\begin{equation}
		I(z) = \delta\lambda(z) \left( \frac{dT^{(0)}(z)}{dz} \right)^2 + i\omega \delta(\rho c_p)(z) \left( T^{(0)}(z) \right)^2
		\label{eq:simplified_integrand}
	\end{equation}
	
	Over the evaluated spatial domain, the hygrothermal degradation is mapped according to its exact macroscopic physical distribution. While the sensible heat capacity increases linearly with the quasi-static moisture gradient (law of mixtures), the effective thermal conductivity scales exponentially (ISO 10456). The continuous spatial discrepancies are therefore strictly defined by a linear capacity gradient $d_1 = \Delta(\rho c_p) / e_j$ and an exponential conductivity growth factor $\beta = \frac{1}{e_j} \ln \left( \lambda(e_j)/\lambda_0 \right)$. This yields the exact local spatial perturbations:
	\begin{align}
		\delta(\rho c_p)(z) &= d_1 z \\
		\delta\lambda(z) &= \lambda_0 \left( e^{\beta z} - 1 \right)
	\end{align}
	
	The zero-order temperature field $T^{(0)}(z)$, propagating through the nominal baseline, is a linear combination of forward and backward thermal waves governed by the nominal complex wave vector $q_j$:
	\begin{equation}
		T^{(0)}(z) = A e^{q_j z} + B e^{-q_j z}
	\end{equation}
	where the complex amplitudes $A$ and $B$ are strictly determined by the boundary conditions established during the zero-order admittance mapping. 
	
	Squaring the temperature field and its spatial derivative yields:
	\begin{align}
		\left( T^{(0)}(z) \right)^2 &= A^2 e^{2q_j z} + B^2 e^{-2q_j z} + 2AB \\
		\left( \frac{dT^{(0)}(z)}{dz} \right)^2 &= q_j^2 \left( A^2 e^{2q_j z} + B^2 e^{-2q_j z} - 2AB \right)
	\end{align}
	
	Substituting these expanded squares and the exact spatial profiles into Eq.~\eqref{eq:simplified_integrand} algebraically separates the integrand into a purely exponential resistive term $I_{res}(z)$ and a polynomial-exponential capacitive term $I_{cap}(z)$:
	\begin{align}
		I_{res}(z) &= \lambda_0 q_j^2 \left[ A^2 \left( e^{(\beta + 2q_j)z} - e^{2q_j z} \right) + B^2 \left( e^{(\beta - 2q_j)z} - e^{-2q_j z} \right) - 2AB \left( e^{\beta z} - 1 \right) \right] \\
		I_{cap}(z) &= z \left[ D_1 e^{2q_j z} + D_2 e^{-2q_j z} + D_3 \right]
	\end{align}
	where the capacitive polynomial coefficients are defined as $D_1 = i\omega d_1 A^2$, $D_2 = i\omega d_1 B^2$, and $D_3 = i\omega d_1 2AB$.
	
	The macroscopic integral $J_j = \int_0^{e_j} \left[ I_{res}(z) + I_{cap}(z) \right] dz$ is thus resolved without resorting to geometric chord approximations. The capacitive component retains its integration by parts, while the exact exponential resistive component integrates directly, yielding the sum of two exact closed-form primitives, $J_j = J_{res} + J_{cap}$:
	\begin{align}
		J_{res} &= \lambda_0 q_j^2 \left[ A^2 \left( \frac{e^{(\beta + 2q_j)e_j}-1}{\beta + 2q_j} - \frac{e^{2q_j e_j}-1}{2q_j} \right) + B^2 \left( \frac{e^{(\beta - 2q_j)e_j}-1}{\beta - 2q_j} - \frac{e^{-2q_j e_j}-1}{-2q_j} \right) \right. \nonumber \\
		&\qquad \left. - 2AB \left( \frac{e^{\beta e_j}-1}{\beta} - e_j \right) \right] \\
		J_{cap} &= D_1 \left[ \left( \frac{e_j}{2q_j} - \frac{1}{4q_j^2} \right) e^{2q_j e_j} + \frac{1}{4q_j^2} \right] \nonumber \\
		&\qquad + D_2 \left[ \left( \frac{e_j}{-2q_j} - \frac{1}{4q_j^2} \right) e^{-2q_j e_j} + \frac{1}{4q_j^2} \right] + D_3 \frac{e_j^2}{2}
	\end{align}
	
	Direct numerical evaluation of this combined primitive using standard floating-point arithmetic remains susceptible to overflow. The expansion generates positive exponential terms (e.g., $e^{2q_j e_j}$ and $e^{\beta e_j}$) which can exceed the 64-bit precision limit in highly dissipative media. To ensure unconditional numerical stability, the continuous temperature field is reformulated using bounded local amplitudes, $\tilde{A}_j$ and $\tilde{B}_j$, defined relative to both boundaries through a local coordinate $z \in [0, e_j]$:
	\begin{equation}
		T^{(0)}(z) = \tilde{A}_j e^{-q_j(e_j - z)} + \tilde{B}_j e^{-q_j z}
	\end{equation}
	
	By substituting this bounded formulation into the expanded primitive, the definite integral across the stratum thickness is evaluated exclusively using negative and bounded exponential arguments, yielding a numerically stable spatial integral $J_j$:
	\begin{equation}
		J_j = \mathcal{F}_{closed}\left( \tilde{A}_j, \tilde{B}_j, e^{-q_j e_j}, e^{-2q_j e_j}, e^{-\beta e_j} \right)
	\end{equation}
	where $\mathcal{F}_{closed}$ represents the algebraically expanded finite evaluation. The stabilized recursive update for the admittance perturbation, propagating outward from the internal interface, is then computed as:
	\begin{equation}
		Y^{(1)}(x_j) = g_j^2 Y^{(1)}(x_{j-1}) + \frac{1}{\left(T^{(0)}(x_j)\right)^2} J_j
	\end{equation}

	For the fundamental stationary component ($\omega = 0$), the complex wave vector and capacitive terms vanish ($q_j \to 0$, $D \to 0$). The macroscopic thermal resistance of the degraded stratum is resolved analytically by direct integration of the exact exponential spatial conductivity profile $\lambda(z)$, bypassing the perturbative Riccati update:
	\begin{equation}
		R_{exact,j} = \int_{0}^{e_j} \frac{dz}{\lambda_0 e^{\beta z}} = \frac{1 - e^{-\beta e_j}}{\beta \lambda_0}
	\end{equation}
	
	This exact closed-form evaluation fundamentally replaces the standard algorithmic requirement for spatial discretization. It directly maps the continuous coupled heat and moisture degradation into the corrected boundary admittance $Y^{(1)}(x_j)$ in a single $\mathcal{O}(1)$ computational step per layer, unconditionally preventing grid-induced diffusion and structural staircase approximations.
	
	\subsection{Transient Impact of the Perturbative Correction}
	
\begin{figure}[htbp]
		\centering
		\includegraphics[width=0.95\textwidth]{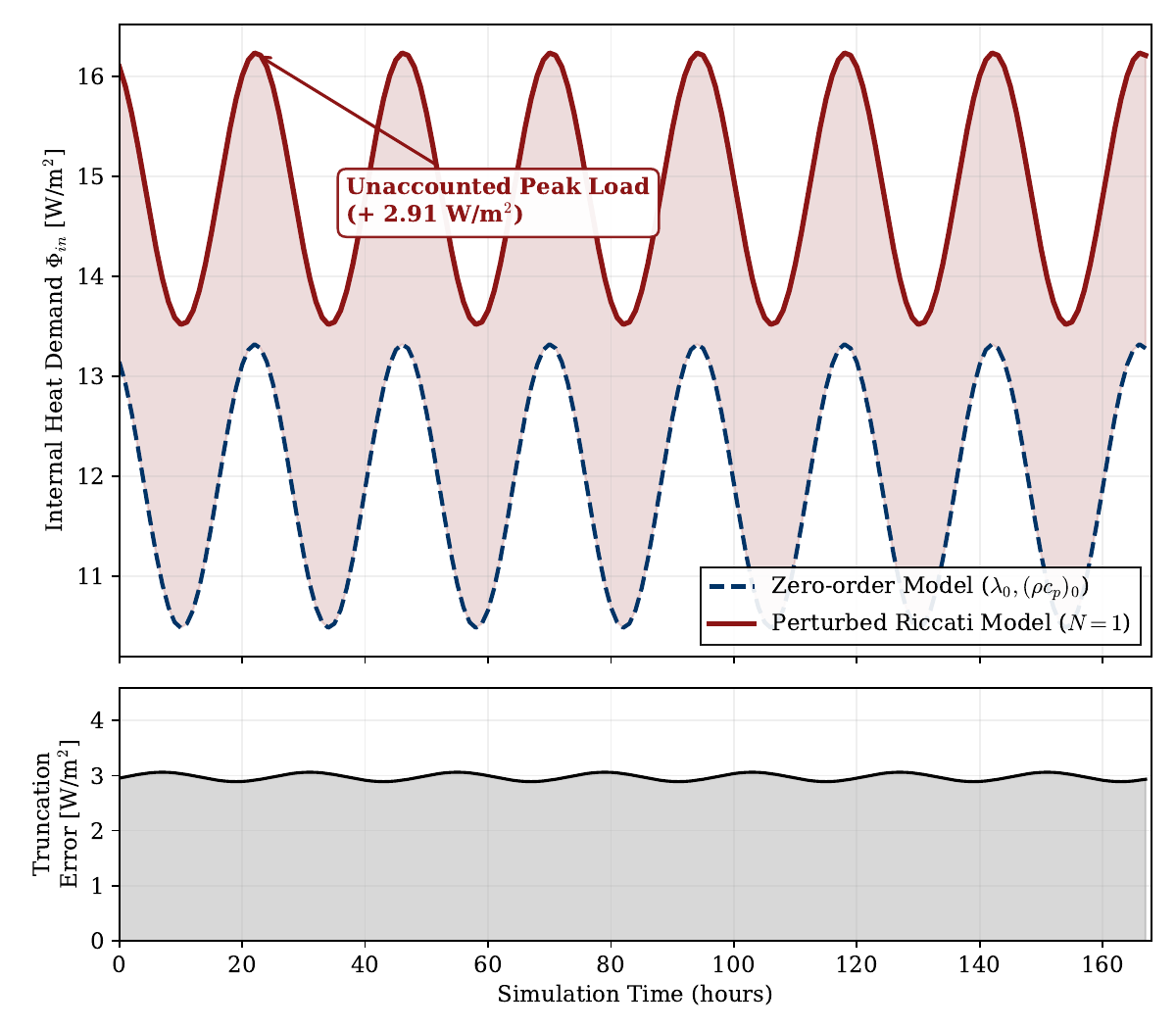}
		\caption{Transient impact of the perturbative correction on the internal heat demand assessment during a 7-day winter sequence. \textbf{(A)} The zero-order model (dashed line) evaluates the thermal admittance using a constant baseline $\lambda_0$ and $(\rho c_p)_0$ (corresponding to a uniform 2\% moisture content), systematically underestimating heat losses. The perturbed model (solid line) dynamically integrates the exponential moisture-driven degradation of the AAC layer, highlighting an unaccounted peak load of $+2.91~\mathrm{W/m}^2$ (a $\sim 21.9\%$ relative deviation) during nocturnal thermal stress. \textbf{(B)} The isolated transient truncation error caused by the homogeneous assumption, addressed by the $\mathcal{O}(N)$ perturbative propagator.}
		\label{fig:perturbative_impact}
	\end{figure}

	The exact analytical closure of the macroscopic primitive mathematically validates the perturbative framework, yet its tangible value lies in the correction of macroscopic boundary fluxes. To quantify the impact of the standard homogeneous assumption, Figure~\ref{fig:perturbative_impact} evaluates the transient internal heat demand ($\Phi_{in}$) of a highly porous 20~cm AAC envelope during a 7-day winter sequence. The external forcing consists of a stochastic cold front (mean temperature of $-2\,^{\circ}\mathrm{C}$ with diurnal oscillations), while the internal boundary is maintained at a $20\,^{\circ}\mathrm{C}$ setpoint.
	
	Under these winter conditions, interstitial condensation typically generates a quasi-static moisture gradient, ranging from a relatively dry internal state (2\% volumetric moisture) to a wetted external boundary (15\% moisture). Relying on the zero-order homogeneous model, evaluated at the nominal interior baseline ($\lambda_0$ and $(\rho c_p)_0$), tends to overestimate the global thermal resistance of the wall by omitting the replacement of insulating pore air with conductive liquid water. 
	
	When this moisture-dependent spatial degradation is integrated via the closed-form analytical resolution of the perturbative propagator $Y^{(1)}(x_j)$, the predicted internal heat flux increases. As demonstrated in Panel A, during periods of nocturnal thermal stress, the analytical correction captures an unaccounted peak heating load of $+2.91~\mathrm{W/m}^2$. For this specific architectural configuration, this represents a relative difference approaching $21.9\%$ on the dynamic thermal load. 
	
	Panel B isolates this transient truncation error over the simulation horizon. For Model Predictive Control (MPC) applications targeting energy minimization or HVAC sizing, a systematic drift of this magnitude can lead to under-heating and compromised thermal comfort. By evaluating the analytical primitive $F_j(\xi)$, the proposed perturbative framework mitigates this structural error. It captures the macroscopic non-linear thermodynamic response of the wetted media without resorting to internal nodal discretization, thereby preserving the $\mathcal{O}(N)$ computational complexity suitable for real-time optimization.
	
	\subsection{Computational Benchmarking and Structural Simplicity}
	
	To formalize the computational characteristics of the continuous admittance mapping, a numerical benchmark was conducted against the standard discretized Transfer Matrix Method (TMM). The absolute ground truth for the dynamic admittance of the wetted AAC layer was defined using a numerical asymptote ($M = 10,000$ discrete nodes, $dx \to 0$) to eliminate spatial truncation. Both the discrete TMM evaluated at practical mesh sizes ($M_s \in [2, 1000]$) and the single-step analytical Riccati propagator were compared against this baseline.
	
	\begin{figure}[htbp]
		\centering
		\includegraphics[width=0.95\textwidth]{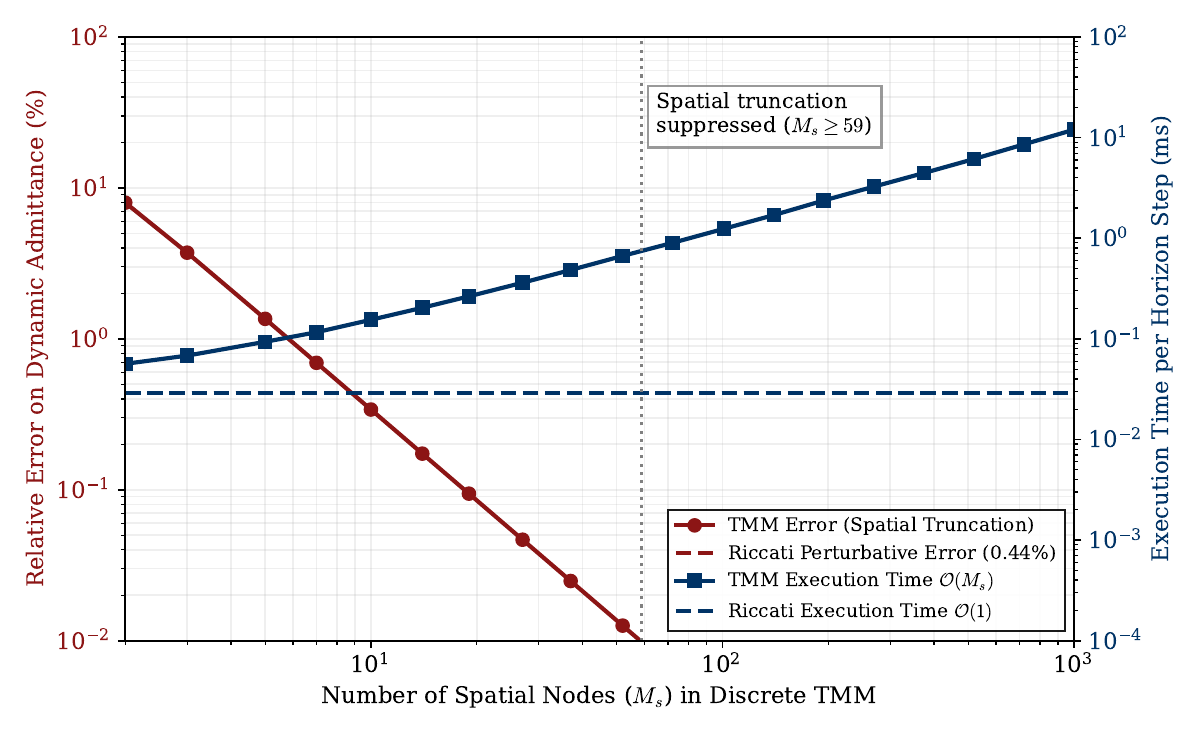}
		\caption{Computational benchmarking of the discrete Transfer Matrix Method (TMM) versus the continuous Riccati admittance framework under a severe hygrothermal gradient. The left axis evaluates the relative spatial truncation error against a continuous numerical asymptote ($M=10,000$). The discrete TMM error decreases quadratically, crossing the $0.44\%$ intrinsic error bound of the first-order Riccati approximation at $M_s \approx 10$. The right axis tracks the execution time per predictive step. Fully suppressing this piecewise structural truncation error in the discrete TMM requires $M_s > 59$ spatial nodes, imposing a strictly linear $\mathcal{O}(M_s)$ computational penalty. Conversely, the Riccati formulation integrates the macroscopic domain analytically in a single $\mathcal{O}(1)$ operation, maintaining constant execution times in the microsecond range ($\approx 0.02~\mathrm{ms}$).}
		\label{fig:benchmark_riccati}
	\end{figure}
	
	Figure~\ref{fig:benchmark_riccati} illustrates this computational duality. The error of the TMM decreases quadratically as the mesh is refined, crossing the $0.44\%$ intrinsic threshold of the Riccati first-order approximation at exactly $M_s = 10$. While enforcing high-density grids ($M_s \ge 59$) allows the discrete method to asymptotically surpass the perturbative fidelity, it fundamentally imposes a strictly linear $\mathcal{O}(M_s)$ execution penalty. Conversely, the Riccati formulation analytically integrates the macroscopic domain in a single $\mathcal{O}(1)$ operation, maintaining constant execution times in the microsecond range ($\approx 0.02~\mathrm{ms}$).
	
	To achieve its accuracy, the discrete TMM requires an algorithmic loop scaling linearly with the nodal density ($\mathcal{O}(M_s)$). The execution time depends on the engineer's ability to carefully select an adequate spatial grid ($M_s$) that balances truncation error and computational overhead for every specific material configuration. In contrast, the Riccati formulation integrates the physical gradients analytically into a closed-form primitive. The macroscopic layer is evaluated in a single deterministic step ($\mathcal{O}(1)$), yielding execution times in the microsecond range ($\approx 0.02~\mathrm{ms}$) regardless of the material's thermal thickness or the severity of the gradient. It completely absolves the solver from grid-dependency and spatial convergence checks.
	
	Beyond the strict modeling of isolated structural components, this meshless reduction dictates the feasibility of system-level energy management. Standard practice often forces engineers to blindly accept auto-meshing defaults (e.g., $M_s \le 10$), which drastically truncates the exponential non-linearities of coupled heat and moisture transfer \cite{tariku2010transient}. Enforcing high-fidelity spatial meshes ($M_s \ge 59$) drastically increases the computational overhead, paralyzing real-time optimization solvers \cite{picard2015impact}.
	
	To illustrate this bottleneck at the urban scale, consider a stochastic Model Predictive Control (SMPC) dispatch optimizing a smart grid of 500 interconnected buildings (comprising 5,000 macroscopic envelopes). Urban Building Energy Modeling (UBEM) demands the simultaneous dynamic resolution of these interconnected elements \cite{reinhart2016urban}. To account for weather and occupancy uncertainties, the SMPC typically executes robust probabilistic scenarios, such as 1,000 Monte Carlo iterations per predictive cycle \cite{drgona2020all}, requiring $5 \times 10^6$ total envelope evaluations. 
	
	Using a fully discretized approach ($M_s=59$) to accurately capture material degradation demands approximately $0.20~\mathrm{ms}$ per evaluation on standard hardware, resulting in a systemic execution time exceeding 16 minutes. This computational lag renders the optimization structurally obsolete for standard 15-minute intraday electricity markets or rapid Demand Response events \cite{deconinck2016practical}. By shifting the resolution of severe continuous physical non-linearities from an iterative spatial matrix cascade ($\mathcal{O}(M_s)$) to the pre-integrated Riccati analytical primitive ($\mathcal{O}(1)$ evaluated in $\approx 0.02~\mathrm{ms}$), the framework processes the identical probabilistic horizon in under 2 minutes. 
	
	While Reduced Order Models (ROMs), such as equivalent Resistance-Capacitance (RC) networks, routinely achieve comparable execution speeds ($\mathcal{O}(1)$) suitable for MPC, they operate inherently as gray-box or lumped-parameter representations. Capturing severe spatial non-linearities, such as transient moisture gradients, within an RC framework requires either extensive data-driven calibration or computationally expensive mathematical reduction techniques (Model Order Reduction) applied to a finely meshed baseline. The proposed Riccati propagator circumvents this calibration bottleneck. It retains the direct physical causality and exactness of a white-box analytical solver while operating at the execution speed of a reduced-order model.
	
	\subsection{Radiative Boundary Conditions as a Harmonic Source Term}
	
	External boundary conditions involve non-linearities, primarily driven by the longwave radiative exchange between the envelope surface and the celestial vault. According to the Stefan-Boltzmann law, the net radiative heat flux is proportional to the fourth power of the thermodynamic temperatures ($q_{lw} \propto T_{surf}^4 - T_{sky}^4$) \cite{stephenson1967cooling}. 
	
	Standard frequency-domain models often circumvent this non-linearity by linearizing the exchange using a constant radiative heat transfer coefficient ($h_{rad}$). However, this assumption tends to underestimate radiative cooling events, such as nocturnal subcooling under clear skies \cite{bliss1961atmospheric, swinbank1963long, martin1984characteristics, berdahl1984emissivity}. Time-domain solvers resolve the $T^4$ dependency but require iterative nodal updates, increasing computational overhead \cite{meng2020review}.
	
	The proposed perturbative framework enables non-linear boundary evaluations without forfeiting the $\mathcal{O}(N)$ frequency-domain execution speed. Instead of modifying the core propagator, the non-linearity is isolated as a flux residual. The zero-order surface temperature $T_{surf}^{(0)}(t)$ is first computed using a linearized baseline. The non-linear radiative flux is then evaluated in the time domain, and the difference relative to the linear baseline defines a time-dependent physical residual $\Delta \Phi_{rad}(t)$.
	Through a discrete Fourier transform, this temporal residual is mapped into a harmonic source term $\Delta \tilde{\Phi}_{rad}(\omega)$. To prevent spectral singularities when mapping this flux onto a boundary operator, specifically at harmonics where the zero-order temperature excitation approaches the numerical noise floor ($|\tilde{T}_{surf}^{(0)}(\omega)| \to 0$), a spectral thresholding protocol is enforced. The recursive admittance mapping is initialized at the external interface ($N$-th layer) using an equivalent frequency-dependent pseudo-admittance:
	\begin{equation}
		Y_N^{(1)}(\omega) = 
		\begin{cases} 
			\frac{\Delta \tilde{\Phi}_{rad}(\omega)}{\tilde{T}_{surf}^{(0)}(\omega)} & \text{for } |\tilde{T}_{surf}^{(0)}(\omega)| \ge \tau_{noise} \\
			0 & \text{for } |\tilde{T}_{surf}^{(0)}(\omega)| < \tau_{noise}
		\end{cases}
	\end{equation}
	where $\tau_{noise}$ represents the absolute spectral amplitude threshold. While standard 64-bit floating-point architectures naturally evaluate minor harmonics down to machine precision ($\mathcal{O}(10^{-16})$), propagating ultra-low energy frequencies through the recursive admittance framework incurs unnecessary algorithmic overhead without contributing to macroscopic thermal dynamics. To strictly isolate physically meaningful thermodynamic excitations from negligible spectral energy, this study enforces a conservative threshold of $\tau_{noise} = 10^{-6}$~K. Harmonics below this macroscopic amplitude carry zero thermodynamic relevance and are systematically bypassed. This active truncation prevents the injection of artificial poles into the Riccati propagator, strictly preserving the numerical conditioning of the global structural matrix.
	
	This non-iterative projection inherently introduces a secondary boundary error. By evaluating the radiative residual based on the warmer unperturbed surface temperature $T_{surf}^{(0)}$, the method bypasses the coupled $T^4$ feedback loop, generating a local surface flux discrepancy of $\mathcal{O}(\Delta T_{surf})$.
	
	For Model Predictive Control (MPC) applications, the target variable is often the transmitted internal heat demand $\Phi_{in}(t)$. The transmission of this secondary surface error through the envelope is governed by the global thermal transfer function $G(\omega)$ (Eq.~\eqref{eq:closure_final}). 
	
	Because multi-layered building envelopes act as low-pass filters with high thermal thicknesses ($\text{Re}(q_j e_j) \gg 1$), external boundary feedback discrepancies are exponentially attenuated. Even if the single-pass assumption yields an uncorrected surface overcooling of $1\,^{\circ}\mathrm{C}$ to $2\,^{\circ}\mathrm{C}$ during peak nocturnal subcooling, the spatial Riccati propagator dampens this secondary discrepancy to a negligible margin (e.g., $< 0.1~\mathrm{W/m}^2$) at the internal interface. While this open-loop boundary formulation is structurally insufficient for local exterior diagnostics (e.g., surface condensation or frost risk assessment), it is perfectly suited for predictive algorithms focused strictly on internal heat demand ($\Phi_{in}$) and indoor thermal comfort mapping.
	
	The primary error generated by the standard LTI linearization can be more significant, especially for moisture-degraded envelopes. As demonstrated in the previous section, the accumulation of interstitial moisture increases the effective thermal conductivity of the AAC layer. Any uncorrected external flux generated by a linearized boundary condition will influence the interior node more readily than it would through a dry insulator.
	
\begin{figure}[htbp]
		\centering
		\includegraphics[width=0.95\textwidth]{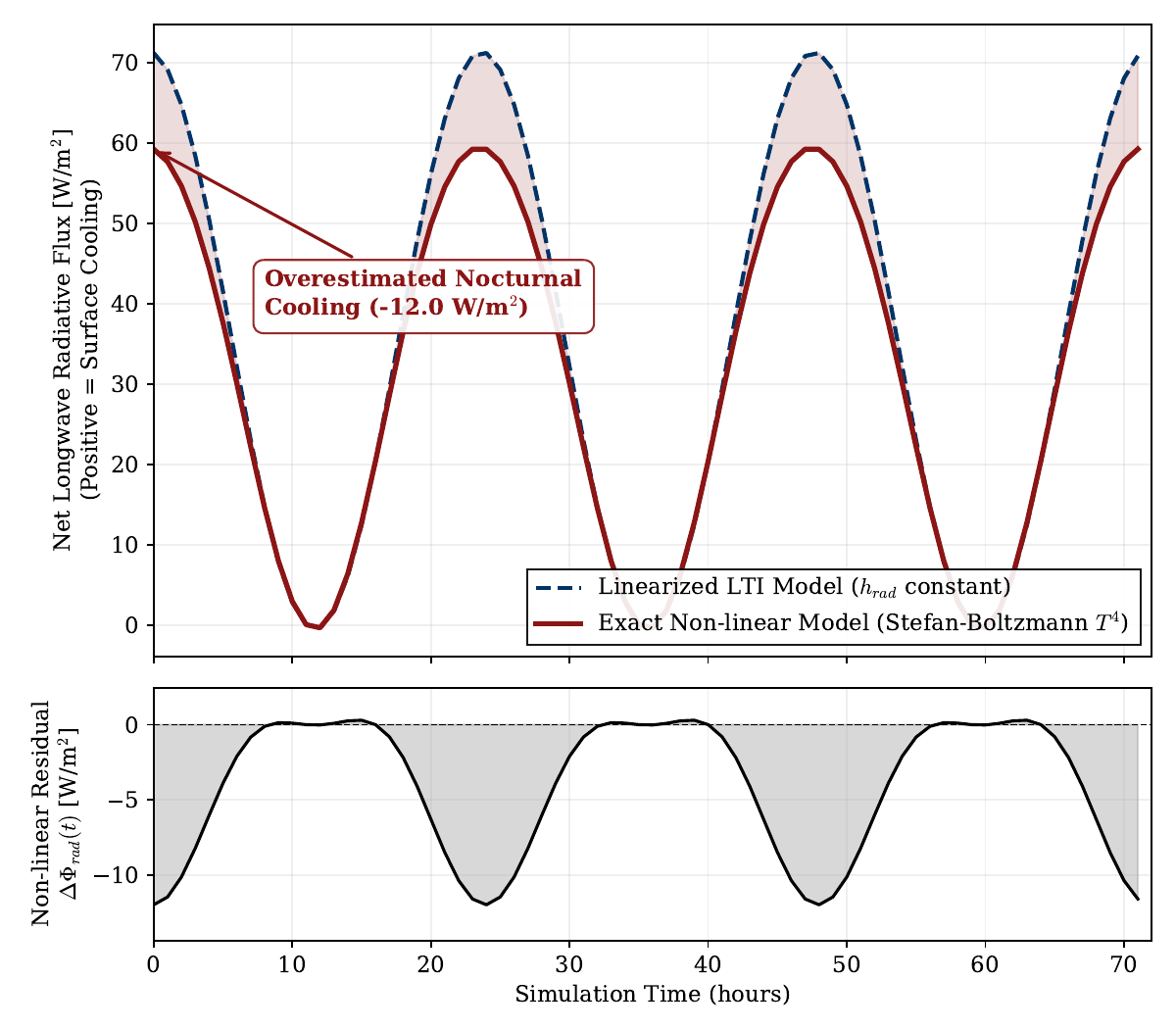}
		\caption{Numerical application of the non-linear radiative boundary condition during a clear-sky winter nocturnal subcooling event. \textbf{(A)} Comparison between the exact Stefan-Boltzmann longwave emission and the standard linearized LTI model evaluated at the mean winter temperature. The linearization overestimates the nocturnal heat discharge, generating an artificial cooling flux of $-12.0~\mathrm{W/m}^2$. \textbf{(B)} The isolated non-linear physical residual $\Delta \Phi_{rad}(t)$. This array is transformed into the frequency domain ($\Delta \tilde{\Phi}_{rad}(\omega)$) to initialize the perturbative admittance $Y_N^{(1)}(\omega)$. The secondary error generated by the single-pass evaluation is exponentially attenuated by the envelope, allowing recovery of the internal boundary thermodynamics without iterative nodal resolution.}
		\label{fig:radiative_impact}
	\end{figure}

	To quantify the influence of this non-linear boundary assessment, Figure~\ref{fig:radiative_impact} evaluates the net longwave radiative exchange during a 3-day clear-sky nocturnal subcooling event, aligned with the winter sequence ($-2\,^{\circ}\mathrm{C}$ average). Under these meteorological conditions, the apparent sky temperature ($T_{sky}$) drops below the ambient air temperature \cite{cole1976longwave}.
	
	The standard LTI assumption fails to capture the thermodynamic behavior of the Stefan-Boltzmann law since it relies on a linearized and constant radiative heat transfer coefficient ($h_{rad}$) evaluated at the mean winter temperature. During peak nocturnal cooling, the linearized gradient overestimates the radiative heat discharge by $12.0~\mathrm{W/m}^2$. In an MPC context managing a conductive wetted envelope, this unaccounted heat loss can inflate the predicted heating demand for the following morning, affecting the optimal energy schedule.
	
	The time-domain difference between the $T^4$ emission and the linear baseline constitutes the primary non-linear physical residual $\Delta \Phi_{rad}(t)$ (Panel B). By spectrally transforming this array and injecting it directly as the perturbed boundary admittance $Y_N^{(1)}(\omega)$, the proposed mathematical framework mitigates this transient overestimation. The attenuation of the envelope absorbs the secondary single-pass error, offering a method to recover the boundary thermodynamics without iterative time-stepping solvers.
	
	\section{Conclusion}
	
	This paper presented a frequency-domain framework based on the continuous spatial Riccati equation to resolve one-dimensional transient heat conduction in multi-layered building envelopes. The derived recursive admittance mapping algebraically bounds positive exponential wave arguments, neutralizing the 64-bit floating-point overflow inherent to standard Transfer Matrix Methods (TMM) under high-frequency dynamic forcing \cite{davies1973thermal}. By integrating spatial derivatives analytically, the formulation eliminates the requirement for internal discretization grids, strictly bypassing Courant-Friedrichs-Lewy (CFL) stability constraints \cite{courant1967partial}. The spatial algorithmic complexity is strictly preserved at $\mathcal{O}(N)$ per harmonic. This structural bypass of state-space inflation enables full time-domain reconstructions ($\mathcal{O}(M \log M + N \cdot M)$) with execution speeds compatible with the stochastic multi-week predictive horizons of Model Predictive Control (MPC) architectures \cite{kansal2025review}.
	
	Time-domain reconstruction via the Inverse Discrete Fourier Transform was stabilized by bounding the temporal aliasing artifact through exponential temporal padding, dictated by the macroscopic thermal relaxation time of the system. To address the physical limitations of the standard Linear Time-Invariant (LTI) homogeneous assumption, the propagator was extended using regular perturbation theory. Continuous spatial gradients (in particular the moisture-driven degradation of structural thermal conductivity ($\lambda(x)$) \cite{elassaad2024influence}) were analytically integrated as equivalent harmonic source terms. For wetted porous media, this closed-form perturbative integration recovered a peak heating load truncation error of $+2.91~\mathrm{W/m}^2$ ($\approx 21.9\%$ relative deviation) without introducing spatial nodal meshes.
	
	The non-linear Stefan-Boltzmann radiative boundary condition ($T^4$) during clear-sky nocturnal subcooling was evaluated as a time-domain residual and mapped into a dynamic boundary admittance correction. This protocol corrected an overestimated nocturnal cooling flux of $-12.0~\mathrm{W/m}^2$ generated by standard linearized boundary conditions \cite{koch2025multi}. The spatial Riccati formulation establishes a mathematically bounded propagator that reconciles the $\mathcal{O}(N)$ execution speed of spectral methods with the non-linear thermodynamic fidelity conventionally restricted to explicit spatially discretized solvers.
	
	\section*{Data availability statement}
	The complete Python source code required to reproduce the numerical benchmarking, the spatial convergence analysis, and all associated figures is openly available in the Zenodo repository at \url{https://doi.org/10.5281/zenodo.19783248}. The repository includes a comprehensive \texttt{README.md} execution guide and relies strictly on standard open-source scientific libraries.
	
	\section*{Conflict of interest}
	The author declares no conflicts of interest.
	
	\section*{Acknowledgments}
	The author acknowledges the use of a Large Language Model (Gemini) for linguistic refinement, proofreading assistance, and LaTeX formatting during the preparation of this manuscript.

	\bibliographystyle{tfnlm} 
	\bibliography{bibliography}
	
\end{document}